\documentclass[usenatbib,useAMS]{mnras}
\usepackage{graphicx}	
\usepackage{amsmath}	
\usepackage{amssymb}	
\usepackage{multicol}        
\usepackage{bm}		
\usepackage{pdflscape}	
\usepackage[T1]{fontenc}
\usepackage{ae,aecompl}
\usepackage{newtxtext,newtxmath}
\usepackage{caption}
\usepackage{threeparttable}
\usepackage{appendix}
\usepackage{soul, color, xcolor}

\title[Kpc-scale dual-core system]{The orbital kinematics of kpc-scale dual-core systems with double-peaked narrow emission 
lines}

\author[X. Q. Chen]{XingQian Chen$^{1}$, Peizhen Cheng$^{1}$, Ying Gu$^{1}$, Qi Zheng$^{2}$, MingFeng Liu$^{2}$, XueGuang Zhang$^{1}$
\thanks{Contact e-mail: \href{mailto:xgzhang@gxu.edu.cn}{xgzhang@gxu.edu.cn}}%
\\
$^{1}$School of Physical Science and Technology, Guangxi University, No. 100, Daxue East Road, Nanning 530004, P. R. China
\\
$^{2}$School of Physics and Technology, Nanjing Normal University, No. 1, Wenyuan Road, Nanjing 210023, P. R. China}
\begin{document}
\bibliographystyle{plain}
\label{firstpage}
\pagerange{\pageref{firstpage}--\pageref{lastpage}}
\maketitle

\begin{abstract}
In this manuscript, based on SDSS photometry and spectroscopy, a method is proposed to test the hypothesis that the orbital kinematics 
of kpc-scale dual-core systems can lead to double-peaked narrow emission lines (DPNELs), through analyzing a sample of seven kpc-scale dual-core 
systems by comparing the upper limits of their orbital velocities (calculated using total stellar mass and projected distance) 
with the velocity separation of DPNELs (peak separation).
To determine accurate total stellar masses, \textit{GALFIT} is applied to consider the effects of the overlapping components on the photometric 
images to obtain accurate magnitudes.
Then, based on the correlation between absolute Petrosian magnitudes and total stellar masses, the individual masses of the galaxies 
are determined.
Therefore, the maximum orbital velocities can be calculated by combining the projected distance. 
Meanwhile, the peak separation can be accurately measured after subtracting the pPXF determined host galaxy contributions. 
Finally, four objects exhibit peak separations almost consistent with their respective maximum orbital velocities under the 
assumption of a circular orbit, while the remaining three objects display peak separations larger than the maximum orbital velocities. 
A larger sample will be given later to further test whether DPNELs can arise from kpc-scale dual-core systems.
\end{abstract}


\section{Introduction}
Dual-core systems are commonly considered to be natural products of galaxy merging. This merging process is also a fundamental mechanism 
in the hierarchical framework of galaxy formation and evolution, as well discussed in \cite{1988ApJ...325...74S, 1991ApJ...379...52W, 
1998A&A...331L...1S, 2000MNRAS.319..168C, 2005Natur.435..629S, 2019MNRAS.490.5390B, 2022ApJ...929..167M}. 
Given that supermassive black holes are found in the centers of almost all galaxies \citep{1995ARA&A..33..581K, 2021ApJ...913..102W}, 
two massive black holes merging at kiloparsec (kpc) scales would lead to the merging remnants of the two galaxies being detected as 
dual-core systems, such as dual active galactic nuclei (dual-AGN) and dual supermassive black holes (dual-SMBH), as well discussed in 
\cite{2012ApJ...753...42C, 2015ApJ...799...72F, 2023MNRAS.519.5149D}. 
Researching on dual-core systems can provide further insights into the final stage of dynamical friction in the merging process and can 
constrain galaxy evolutionary models \citep{2012ApJ...748L...7V, 2013MNRAS.429.2594B, 2014SSRv..183..189C}.
A kpc-scale dual-core system containing two independent narrow line regions (NLRs) can result in double-peaked narrow emission lines 
(DPNELs) due to orbital motion, making DPNELs potential indicators for dual-core systems.

\cite{2004ApJ...604L..33Z} have confirmed a dual-core system in SDSS J1048+0055 by combinations of double-peaked 
[O~\textsc{iii}]$\lambda$$\lambda$ 4959\AA, 5007\AA\ emission lines (hereafter [O~\textsc{iii}] DPNELs) and radio properties. 
\cite{2007ApJ...660L..23G} have confirmed a dual-core system in EGSD2 J1420+5259 by combinations of [O~\textsc{iii}] DPNELs and 
multi-wavelength photometric image observations. 
\cite{2009ApJ...705L..20X} have confirmed a dual-core system in SDSS J1316+1753 by combinations of all DPNELs and an extra 
broad component between each pair of two narrow emission line systems. 
\cite{2010ApJ...715L..30L} have confirmed four dual-core systems by combinations of [O~\textsc{iii}] DPNELs and deep near-infrared 
(NIR) images. 
\cite{2011ApJ...740L..44F} have confirmed a dual-core system in SDSS J1502+1115 by combinations of [O~\textsc{iii}] DPNELs and 
high-resolution radio images. 
\cite{2011ApJ...738L...2M} have confirmed a dual-core system in SDSS J0952+2552 by combinations of [O~\textsc{iii}] DPNELs and 
NIR integral field spectrograph. 
\cite{2014MNRAS.437...32W} have reported a dual-core system in SDSS J1323-0159 by combinations of double-peaked [O~\textsc{iii}], 
H$\beta$ emission lines and Hubble Space Telescope (HST) imaging. 
\cite{2019ApJ...879L..21G} have confirmed a dual-core system in SDSS J1010+1413 by combinations of HST imaging.
\cite{2021A&A...646A.153S} have reported a dual-core system in SDSS J1431+4358 by combinations of DPNELs and NIR diffraction limited imaging.

However, there are also results suggesting that DPNELs are not reliable indicators for kpc-scale dual-core systems.
\cite{2010ApJ...708..427L} have presented that DPNELs can arise from NLR kinematics. 
\cite{2010ApJ...716..131R} have confirmed that DPNELs can result from jet-driven outflows based on radio image observations. 
\cite{2011ApJ...727...71F} have confirmed that DPNELs can be produced by outflows in the NLR through kinematic model analysis. 
\cite{2011ApJ...735...48S} have reported that a majority of [O~\textsc{iii}] DPNELs can be explained by NLR kinematics in a single AGN. 
\cite{2012ApJ...745...67F} have presented that only $\sim$1\% of dual AGN can result in DPNELs. 
\cite{2015MNRAS.449L..31Z} have reported that double-peaked narrow H$\alpha$ emission features can be explained by non-kinematic models. 
\cite{2018ApJ...854..169L} have suggested that radio-loud DPNELs should be generated by jets. 
\cite{2023MNRAS.520.6190Z} have presented that DPNELs are disfavourably produced by a dual-core system based on flux ratio analysis.

Therefore, a method is proposed to test whether kpc-scale dual-core systems can lead to DPNELs, by comparing the orbital velocity 
of a dual-core system (determined by the total stellar mass and the projected distance) with the velocity separation of the two peaks 
in DPNELs (defined as the velocity shift of the red-shifted component relative to the blue-shifted component in double-peaked profiles, 
hereafter peak separation).
Once the DPNELs originate from a kpc-scale dual-core system, the orbital velocity should not be smaller than the peak separation of 
DPNELs when considering projected effects. 
Based on the assumption of a circular orbit and point-like galaxies, the orbital velocity can be described as:
\begin{equation}
V = \sqrt{\frac{G(M_1+M_2)}{D/(sin\theta \times cos\beta)}}
\end{equation}
, where $D$ is the projected distance between the main galaxy (with DPNELs) and the companion galaxy in the system, which can be 
directly measured in high-quality photometric images, such as the results shown in \cite{2012ApJ...753...42C,2018ApJ...862...29L,
2020ApJ...899..154S}; $\theta$ is the inclination angle of the orbital plane and $\beta$ is the orientation angle; $M_1$ and $M_2$ 
are the total stellar masses of the two galaxies.
The calculation is similar to the approaches in \cite{2009ApJ...705L..76W, 2017MNRAS.465.4772R,2024MNRAS.531L..76Z}. 
However, due to the extended masses of galaxies and ambiguous $\theta$ and $\beta$, one can only estimate the maximum orbital 
velocity ($V_{max}$) of a dual-core system by combining the projected distance with stellar masses, which assumes $\theta$=90° and 
$\beta$=0° (sin$\theta$ $\times$ cos$\beta$=1).
Therefore, the determined peak separation of the DPNELs of the main galaxy can be compared with $V_{max}$, to provide clues to support 
or refute the hypothesis that dual-core systems lead to DPNELs. 

Among the reported objects with DPNELs in \cite{2012ApJS..201...31G}, which is based on the multi-band imaging and spectroscopy of 
galaxies from the Sloan Digital Sky Survey (SDSS) DR7, we select seven kpc-scale dual-core systems based on their apparent dual cores 
in the photometric images and separated spectroscopic results of the two galaxies in each system.
At the current stage, host galaxy masses can be conveniently estimated using the following two methods.
On the one hand, the masses can be derived from spectroscopic results such as using penalized pixel-fitting (pPXF) \citep{2017MNRAS.466..798C} 
or principal component analysis (PCA) \citep{2012MNRAS.421..314C,2019ApJ...883...83P}.
On the other hand, the masses can be derived from photometric results such as using spectral energy distribution based on photometric data 
\citep{2014ApJS..210....3M} or mass-magnitude relation \citep{2010ApJ...708..137M}.
Considering the limited fiber diameter of 3$\arcsec$ leading to smaller stellar masses from the spectroscopic results than the 
intrinsic values for low-redshift objects \citep{2005PASP..117..227K}, the estimations from the photometric results are preferred.
However, for kpc-scale dual-core systems, the mass estimations for $M_1$ and $M_2$ are not accurate enough due to the overlapping 
components covered in the photometric images. 
It is important to correct the effects of the overlapping components on  $V_{max}$ through improved mass estimations. 
Therefore, to address this problem, \textit{GALFIT} \citep{2002AJ....124..266P, 2010AJ....139.2097P} is applied to separate the 
overlapping components in the SDSS photometric images to determine the intrinsic photometric properties of each galaxy in a dual-core system.

This manuscript is organized as follows. Section 2 presents the photometric results. Section 3 presents the spectroscopic properties 
of DPNELs.  Section 4 presents the main results. Conclusions are provided in Section 5. The cosmological model adopted in this 
manuscript assumes the following parameters: \( H_0 = 70 \, \text{km s}^{-1} \text{Mpc}^{-1} \), \( \Omega_m = 0.3 \), 
and \( \Omega_\Lambda = 0.7 \).

\begin{figure*}
  \centering
  \includegraphics[width=0.88\paperwidth]{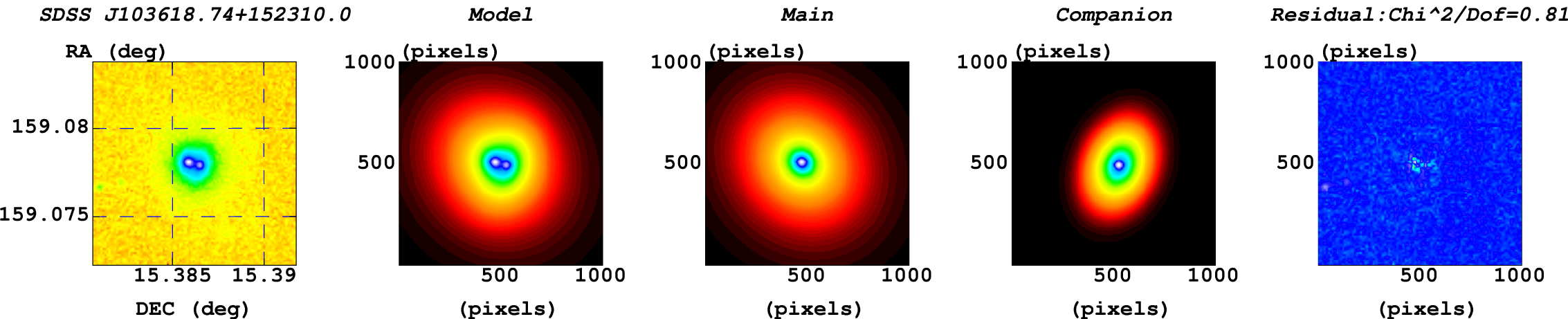}
  \caption{The best fitting results by \textit{GALFIT} to the $r$ band photometric image of the dual-core system in 
SDSS J103618.74+152310.0. 
Column 1: The image region cut from the FITS (Flexible Image Transport System) image of the SDSS field, with dashed lines 
representing the coordinates.
Column 2: The best fitting results. 
Column 3: The separated image of the main galaxy determined by \textit{GALFIT}. 
Column 4: The separated image of the companion galaxy determined by \textit{GALFIT}. 
Column 5: The residual image with the corresponding $\chi2/dof$ marked in the title.
All images are expanded from 101$\times$101 pixels to 1010$\times$1010 pixels using linear interpolation for enhanced visual clarity.}
  \label{GALFIT}
\end{figure*}

\begin{figure*}
 \includegraphics[width=0.85\paperwidth]{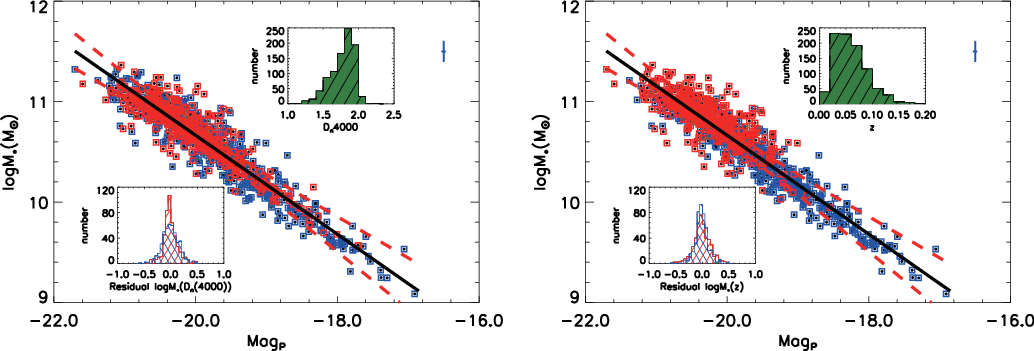} 
 \centering
 \caption{
On the correlation between logarithmic total stellar mass using model photometry from galSpecExtra
and $g$ band absolute Petrosian magnitude for the collected 901 elliptical galaxies.
The solid lines in black represent the best fitting results, and the dashed lines in red show the corresponding 5$\sigma$ confidence bands. 
The standard error bar is shown in the top right region of each panel, determined by averaging the uncertainties 
of logarithmic total stellar masses and absolute Petrosian magnitudes of the 901 elliptical galaxies provided by SDSS.
In the left panel, the black points with red/blue squares represent the results with $D_{n}(4000)$ larger/smaller than the 
median value (1.81); the distribution of $D_{n}$(4000) is shown in the top right region; the histograms filled with red/blue 
represent the distribution of $logM_{*}$($D_{n}$(4000))=$log{M_{*}}$-$\alpha$$\times$$Mag_{P}$-$\beta$ for the results with larger/smaller 
$D_{n}$(4000), shown in the bottom left region.
In the right panel, the black points with red/blue squares represent the results with redshifts larger/smaller than the 
median value (0.05), the distribution of redshift is shown in the top right region; the histograms filled with red/blue 
represent the distribution of $logM_{*}$(z)=$log{M_{*}}$-$\alpha$$\times$$Mag_{P}$-$\beta$ for the results with larger/smaller redshift, 
shown in the bottom left region.
$\alpha$ and $\beta$ are the best fitting results of the mass-magnitude correlation of $g$ band, all residuals are calculated relative to 
this results. 
}
 \label{mass_mag}
\end{figure*}

\begin{table*}
\begin{threeparttable}
 \caption{Main results}
 \label{main_results}
 
 \begin{tabular*}{0.85\paperwidth}{c@{\hspace*{5pt}}c@{\hspace*{5pt}}c@{\hspace*{5pt}}c@{\hspace*{5pt}}c@{\hspace*{5pt}}c@{\hspace*{5pt}}c
@{\hspace*{5pt}}c@{\hspace*{5pt}}c@{\hspace*{5pt}}c@{\hspace*{5pt}}c@{\hspace*{5pt}}c@{\hspace*{5pt}}c@{\hspace*{5pt}}c@{\hspace*{5pt}}c@{\hspace*{5pt}}c@{\hspace*{3pt}}c}
 
 \hline
  Type & SDSS Name & MJD & Plate & Fiber & z & $\triangle V_{peak}$ & $V_{max}^{p}$ & $V_{max}^{c}$ & $D$ & $R$ & $m_{p}$ & $m_{c}$ & $M_{*}^{p}$ & $M_{*}^{c}$ & $IC$ & $D_{n}$(4000)  \\
  (1) & (2) & (3) & (4) & (5) & (6) & (7) & (8) & (9) & (10) & (11) & (12) & (13) & (14) & (15) & (16) & (17)\\
 \hline
 main    & J103618.74+152310.0 & 54177 & 2594 & 244 & 0.066 & 421.9±41.0 & 488.9±103.9 & 315.4±67.1 & 2.25 & 8.59  & 15.87 & 16.39 & 10.97 & 10.59 & 0.25 & 1.84\\
 comp    & J103618.65+152311.9 & 54178 & 2592 & 26  & 0.067 &             &              &             &      & 6.18  & 15.96 & 16.99 & 10.94 & 10.56 & 0.20 & 1.33\\
\hline
 main    & J021242.27+002903.7 & 53763 & 1507 & 421 & 0.150 & 222.6±53.5 & 520.4±103.4 & 336.2±66.8 & 2.63 & 6.81  & 16.61 & 17.58 & 11.48 & 11.03 & 0.22 & 1.55\\
 comp    & J021242.18+002906.1 & 51816 & 405  & 467 & 0.149 &             &              &             &      & 5.11  & 16.71 & 17.32 & 11.43 & 11.12 & 0.16 & 1.85\\
\hline
 main    & J081948.04+254329.0 & 52962 & 1585 & 450 & 0.082 & 158.5±22.1 & 425.6±73.1  & 303.8±52.2 & 3.97 & 3.09  & 16.26 & 17.48 & 11.02 & 10.46 & 0.12 & 1.66\\
 comp    & J081948.22+254330.4 & 52709 & 1266 & 56  & 0.082 &             &              &             &      & 13.27 & 15.71 & 16.01 & 11.30 & 11.10 & 0.27 & 2.02\\
\hline
 main    & J142606.64+202831.5 & 54552 & 2787 & 183 & 0.077 & 279.3±16.7 & 641.1±127.4 & 422.2±83.9 & 2.63 & 3.64  & 15.51 & 16.20 & 11.32 & 11.00 & 0.18 & 1.31\\
 comp    & J142606.51+202829.8 & 54534 & 2774 & 336 & 0.077 &             &              &             &      & 3.92  & 15.51 & 16.35 & 11.33 & 10.92 & 0.16 & 1.33\\
\hline
 main    & J145857.23+090232.3 & 53884 & 1815 & 377 & 0.138 & 207.0±21.2 & 708.2±145.4 & 295.8±60.7 & 2.44 & 4.12  & 16.75 & 17.30 & 11.40 & 11.05 & 0.43 & 1.40\\
 comp    & J145857.08+090232.7 & 54212 & 1715 & 131 & 0.137 &             &              &             &      & 9.46  & 16.14 & 18.26 & 11.81 & 10.65 & 0.31 & 1.56\\
\hline
 main    & J090012.69+183439.1 & 53729 & 2283 & 160 & 0.081 & 273.8±25.2 & 269.7±41.7  & 183.5±28.4 & 5.39 & 4.71  & 16.29 & 16.95 & 11.00 & 10.68 & 0.18 & 1.32\\
 comp    & J090012.30+183436.8 & 53700 & 2285 & 316 & 0.080 &             &              &             &      & 3.68  & 16.47 & 17.21 & 10.90 & 10.54 & 0.24 & 1.47\\
\hline
 main    & J000249.06+004504.8 & 51793 & 388  & 345 & 0.087 & 479.1±14.7 & 455.4±71.7  & 298.9±47.1 & 5.49 & 6.34  & 15.59 & 16.23 & 11.39 & 11.08 & 0.21 & 1.61\\
 comp    & J000249.43+004506.7 & 52203 & 685  & 593 & 0.087 &             &              &             &      & 8.51  & 15.52 & 16.44 & 11.42 & 10.99 & 0.25 & 1.93\\
\hline
 \end{tabular*}
\begin{tablenotes}[flushleft]
\setlength\labelsep{0pt}
\small
\item{Notice: Key parameters of the seven kpc-scale dual-core systems. 
Type shows the component of a dual-core system, main represents the main galaxy with DPNELs, comp represents the 
companion galaxy; 
$\triangle V_{peak}$ is the peak separation in units of km/s; 
$V_{max}^{p}$ is the primary maximum orbital velocity in units of km/s; 
$V_{max}^{c}$ is the corrected maximum orbital velocity in units of km/s; 
$D$ is the projected distance in units of kpc; 
$R$ is the Petrosian radius measured from the $r$ band separated image in units of arcsec; 
$m_{p}$ is the primary apparent Petrosian magnitude measured from the $r$ band original image;
$m_{c}$ is the corrected apparent Petrosian magnitude measured from the $r$ band separated image; 
$M_{*}^{p}$ is the final primary logarithmic total stellar mass in units of $M_{\odot}$; 
$M_{*}^{c}$ is the final corrected logarithmic total stellar mass in units of $M_{\odot}$; 
$IC$ is the inverse concentration index measured from the separated image; 
$D_{n}$(4000) is the 4000\AA -break strength. 
}
\end{tablenotes}
\end{threeparttable}

\end{table*}

\section{PHOTOMETRIC RESULTS}
	 
The sample selected from \citet{2012ApJS..201...31G} in this manuscript includes seven low-redshift narrow emission line galaxies with 
double-peaked features, consisting of five dual-core systems separated by $\leq$ 3$\arcsec$ and two separated by $\geq$ 3$\arcsec$. 
These kpc-scale dual-core systems are selected based on the apparent dual cores (no more than two nuclei) with separated spectra, 
indicating that the two galaxies in each system have individual spectroscopic results.
The spectra of the two galaxies in each dual-core system are used to obtain similar redshifts to ensure that they form a merging system. 
Here, the two dual-core systems separated by $\geq$ 3$\arcsec$ are randomly selected as examples to show that DPNELs in a system with 
a larger projected separation ($\geq$ 3$\arcsec$) are unlikely to be produced by the orbital kinematics, which can be verified through 
the method proposed in the manuscript.

In order to investigate the photometric properties of these kpc-scale dual-core systems, SDSS adopts \textit{Deblender} to separate 
nearby objects in overlapping images.
However, this approach is ineffective and imprecise in separating the overlapping components.
To address this problem, similar to what has been done in \cite{2010AJ....139.2097P, 2019ApJ...880...57M}, \textit{GALFIT} is 
applied to separate the overlapping components and to obtain individual fluxes of the two galaxies in the photometric image. 
Therefore, the magnitude of each galaxy can be accurately measured. 
The method is introduced as follows, along with the measurement of total stellar mass through the mass-magnitude correlation. 

Firstly, during the \textit{GALFIT} fitting procedure, a flat sky model is applied to fit the sky background noise. 
Each galaxy is described with a general S{\'e}rsic profile \citep{1963BAAA....6...41S} along with a point spread function (PSF) in 
each dual-core system. 
Here, the PSF is constructed using the neighboring stars in the photometric image. 
Figure \ref{GALFIT} shows the \textit{GALFIT} determined best fitting results for the $r$ band photometric image of the dual-core 
system in SDSS J103618.74+152310.0, including separate resolved images. 
The same photometric results for the other six dual-core systems are provided in Appendix A. 
Moreover, based on the central positions provided by \textit{GALFIT} for the two galaxies in the photometric image of each dual-core system, 
the projected distance can also be determined, as listed in Column 10 of Table \ref{main_results}.
Furthermore, the decomposition process applied in higher-quality images, such as the DESI image, is also provided in Appendix A.

Secondly, the Petrosian radius and Petrosian magnitude are calculated based on the resolved photometric images. 
The related methodologies are delineated in SDSS algorithms (\url{https://www.sdss4.org/dr17/algorithms/magnitudes/#mag_petro}).
Additionally, PetroR50 and PetroR90 are also measured to determine the inverse concentration index ($IC$, defined as PetroR50/PetroR90) 
to test whether the galaxy is an elliptical galaxy, as explained in \citet{2001AJ....122.1861S, 2001AJ....122.1238S}.

For the main and companion galaxies in each dual-core system, the corrected magnitudes are measured from the separated images 
determined by \textit{GALFIT} (Columns 3 and 4 of Figure \ref{GALFIT}). 
Considering that the Petrosian radius measured from the separated image can better describe the intrinsic photometric properties, 
the radius is also applied to the original image (Column 1 of Figure \ref{GALFIT}) to measure the primary magnitudes for both galaxies, 
to investigate the effects of the overlapping components.
The corresponding Petrosian parameters are listed in Table \ref{main_results}. 

Thirdly, it is generally accepted that there is a scaling relation between total stellar masses and magnitudes for galaxies with different stellar 
populations and different redshifts \citep{2011ApJ...732...12R,2016A&A...585A.160A, 2016ApJ...831...63X, 2021MNRAS.507.5477C, 2021ApJ...912..145N}.  
The mass-magnitude relation is then applied to determine the total stellar mass from the photometric results. 
Given that the $IC$ presented in Table \ref{main_results} indicates that the galaxies in the dual-core systems are mainly elliptical galaxies, 
a dedicated sample including 901 elliptical galaxies is constructed using the Structured Query Language (SQL) search tool 
offered by SDSS DR16 to determine the mass-magnitude relation.  
The selected parameters include the Petrosian magnitude from PhotoObjAll and the total stellar mass using model photometry from galSpecExtra, 
as provided by the SDSS public database. 

Before proceeding further, it is imperative to consider the effects of different initial mass functions (IMF) and stellar evolution models 
on the measured total stellar mass, as discussed in \cite{2009ApJ...699..486C, 2010ApJ...708...58C}.
The selected total stellar mass using model photometry in SDSS is obtained through the method described in \cite{2003MNRAS.341...33K}.  
The IMF adopted from \cite{2001MNRAS.322..231K} is preferred for elliptical galaxies, as discussed in \cite{2006MNRAS.366.1126C,2012MNRAS.421..314C}. 
The stellar evolution model adopted from \cite{2003MNRAS.344.1000B} includes detailed stellar evolution prescriptions presented in 
\cite{2000ApJ...543..644L}.
Since the selected mass-magnitude sample consists of elliptical galaxies, the total stellar mass from galSpecExtra can be reliably 
applied to measure the mass-magnitude relation.
 
The selection criteria for the mass-magnitude sample include three conditions in the $g,r,i$ bands: 
$IC$ is less than 0.35; $fracdeV$ is in the range 0.6 $\sim$ 0.9 
($fracdeV$ represents the weight of the $deV$ component in the combined $deV$ plus $Exp$ model, $deV$ means the de Vaucouleurs model fit, $exp$ means 
the exponential model fit); 
$devab$ is in the range 0.1 $\sim$ 0.5 ($devab$ represents the $a/b$ ratio of the $deV$ fit, with $a$ and $b$ being the  
semi-major and semi-minor axes of the $deV$ fit, respectively). 
After collecting the mass-magnitude sample, the linear relation is estimated using the public \texttt{lts\_linefit} code, 
which can be described as:
\begin{equation}
log{M_{*}} = \alpha\times Mag_{P} + \beta 
\end{equation}
, where $log{M_{*}}$ is the logarithmic total stellar mass and $Mag_{P}$ is the absolute Petrosian magnitude.

The measured mass-magnitude correlations of the $g,r,i$ bands are used to calculate the total stellar mass, as the relations exhibit 
higher confidence levels than those of the $u,z$ bands. 
The rank correlation coefficient ($r_s$) and the best fitting results with their corresponding 1$\sigma$ uncertainties are: 
$r_s$=-0.90, $\alpha$=-0.49±0.01 and $\beta$=0.83±0.16 for the $g$ band; 
$r_s$=-0.93, $\alpha$=-0.47±0.01 and $\beta$=0.88±0.16 for the $r$ band; 
$r_s$=-0.94, $\alpha$=-0.46±0.01 and $\beta$=0.89±0.16 for the $i$ band. 
These results are consistent with the best fitting relation ($\alpha$$\sim$-0.49) for the total stellar mass versus the $r$ band
absolute magnitude reported in \cite{2021MNRAS.507.5477C}.
The final total stellar mass is determined by averaging the measured masses of the three bands. 
Figure \ref{mass_mag} presents the best fitting results for the $g$ band and corresponding 5$\sigma$ confidence bands. 
Table \ref{main_results} lists the primary logarithmic total stellar mass (derived from the primary absolute Petrosian magnitude)  
and the corrected logarithmic total stellar mass (derived from the corrected absolute Petrosian magnitude), 
with detailed uncertainties provided in Appendix B.
Appendix B also includes measured magnitudes and masses for the $g,r,i$ bands.

For the 901 elliptical galaxies of the mass-magnitude sample, to explore the effects of stellar populations with different ages on 
the best fitting results of the mass-magnitude correlation, we select the 4000\AA -break strength ($D_{n}$(4000)) from the SDSS galSpecIndx database 
to trace their stellar ages, as discussed in \cite{2003MNRAS.341...33K}. 
As shown in the left panel of Figure \ref{mass_mag}, through the Student's T-statistic technique, the probability is 
higher than 60\% that the two distributions of $logM_{*}$($D_{n}$(4000))=$log{M_{*}}$-$\alpha$$\times$$Mag_{P}$-$\beta$ for 
the results with larger/smaller $D_{n}$(4000) have the same mean values, strongly indicating few effects of stellar ages on the correlations. 
Meanwhile, the effects of stellar populations with different evolution histories (traced by redshift) are also shown in the right panel 
of Figure \ref{mass_mag}. 
Through the same Student's T-statistic technique, the probability is higher than 60\% that the two distributions of 
$logM_{*}$(z)=$log{M_{*}}$-$\alpha$$\times$$Mag_{P}$-$\beta$ for the results with larger/smaller redshifts have the same mean values, 
strongly indicating few effects of evolution histories on the correlations.

Here, one more point requires clarification.
As shown in the top right panel of Figure 21 in \cite{2003MNRAS.341...33K}, the $D_{n}$(4000) values of the galaxies in the reference are 
evenly distributed from 1.0 to 2.3. 
Meanwhile, the distribution of $D_{n}$(4000) in our mass-magnitude sample is concentrated, with a median value of 1.81 and a standard deviation 
of 0.19, as shown in the top right region of the left panel of Figure \ref{mass_mag}. 
This concentration is the main reason for the weak dependence of total stellar masses on stellar ages in our sample.

In addition, to evaluate the accuracy of the recalculated Petrosian parameters and assess the uniqueness of separating  
the overlapping photometric images into two galaxies by \textit{GALFIT} (which may be affected by degeneracies in the fits image),  
50 single galaxies are randomly selected from the mass-magnitude sample for analysis. 
As detailed in Appendix C, the recalculated apparent Petrosian magnitudes are fully consistent with those 
provided by SDSS, indicating that the decomposition procedure is efficient enough.

Following the three-step analytical process, the overlapping components covered in the photometric images are successfully separated, 
leading to respective Petrosian magnitudes and total stellar masses for each galaxy in each dual-core system, 
both with and without separating the overlapping components.

\section{SPECTROSCOPIC RESULTS}

Based on the separated spectra provided by SDSS for the two galaxies in each kpc-scale dual-core system, the seven objects exhibit 
the same spectral characteristics: the main galaxy shows DPNELs, while the companion galaxy shows normal single-peaked emission features. 
We have also measured the offsets between the drilled positions of SDSS fibers and the \textit{GALFIT} determined central positions 
for the galaxies in the seven systems, as shown in the last column of Table B1 in Appendix B. 
The offsets are all smaller than 0.64$\arcsec$, indicating that the spectra provided by SDSS adequately cover the central regions of the galaxies.

The focus of the manuscript is to analyze the peak separation of DPNELs. 
Although \cite{2012ApJS..201...31G} provided host galaxy contributions for the main galaxy, equivalent analyses for the companion galaxy 
are not presented.
The penalized pixel fitting (pPXF) code \citep{2017MNRAS.466..798C}, an improved Simple Stellar Population (SSP) method based on stellar 
population template spectrum, is adopted to determine the host galaxy contributions and redshifts for both the main and companion galaxies. 
After considering a regularization measurement, this process is conducted through 224 SSPs covering 53 population ages (0.06Gyrs $\sim$ 17.18Gyrs) 
and 12 metallicities (-2.32$\sim$ 0.22).
The redshifts of the two galaxies in each dual-core system, derived through this method and listed in Column 6 of Table \ref{main_results}, 
are sufficiently similar to confirm their merging association.
We have also measured the $D_{n}$(4000) values for the galaxies in these systems using the pPXF code determined host galaxy features, which 
indicates older stellar ages (listed in the last Column of Table \ref{main_results}). 
Notably, the pPXF code is applied to the spectrum of the companion galaxy solely to derive its redshift and $D_{n}$(4000), with no further 
spectroscopic analysis performed on the companion galaxy.

After subtracting the host galaxy contributions determined by the pPXF code from the spectrum of the main galaxy, the narrow emission lines around 
H$\alpha$ (rest wavelength from 6500 to 6800 \AA) and H$\beta$ (rest wavelength from  4800 to 5050 \AA) are mainly considered, including narrow 
Balmer emission lines and [N~\textsc{ii}], [O~\textsc{iii}], [S~\textsc{ii}].
Two Gaussian functions are used to describe each double-peaked profile, and the \texttt{MPFIT} package (a Levenberg-Marquardt least-squares 
minimization technique) is used to fit the DPNELs. 
The peak separation (in units of km/s) of each DPNEL is then measured based on the best determined central wavelengths of double-peaked profiles. 
The calculated peak separations with their corresponding uncertainties for the seven kpc-scale dual-core systems are listed in Column 7 of Table 
\ref{main_results}. 
These uncertainties are derived from the 1$\sigma$ errors of the central wavelengths of the double-peaked features, which are calculated via the 
covariance matrix of the multiple Gaussian functions parameters generated by the \texttt{MPFIT} package.
All the peak separations are similar to the values reported in \cite{2012ApJS..201...31G}.

As an illustrative example, the spectroscopic results of the main and companion galaxies of the dual-core system in SDSS J103618.74+152310.0 are 
presented in Appendix D, along with the host galaxy features derived by the pPXF code.
The best fitting results for the emission lines around H$\alpha$ and H$\beta$ in the main galaxy are also presented.
Furthermore, considering that the manuscript focuses on the orbital kinematics of the dual-core system and the emission line features of the main 
galaxy, whether the companion galaxies are AGNs or star-forming galaxies has little effects on our discussed results.

\begin{figure}
  \includegraphics[width=0.85\columnwidth]{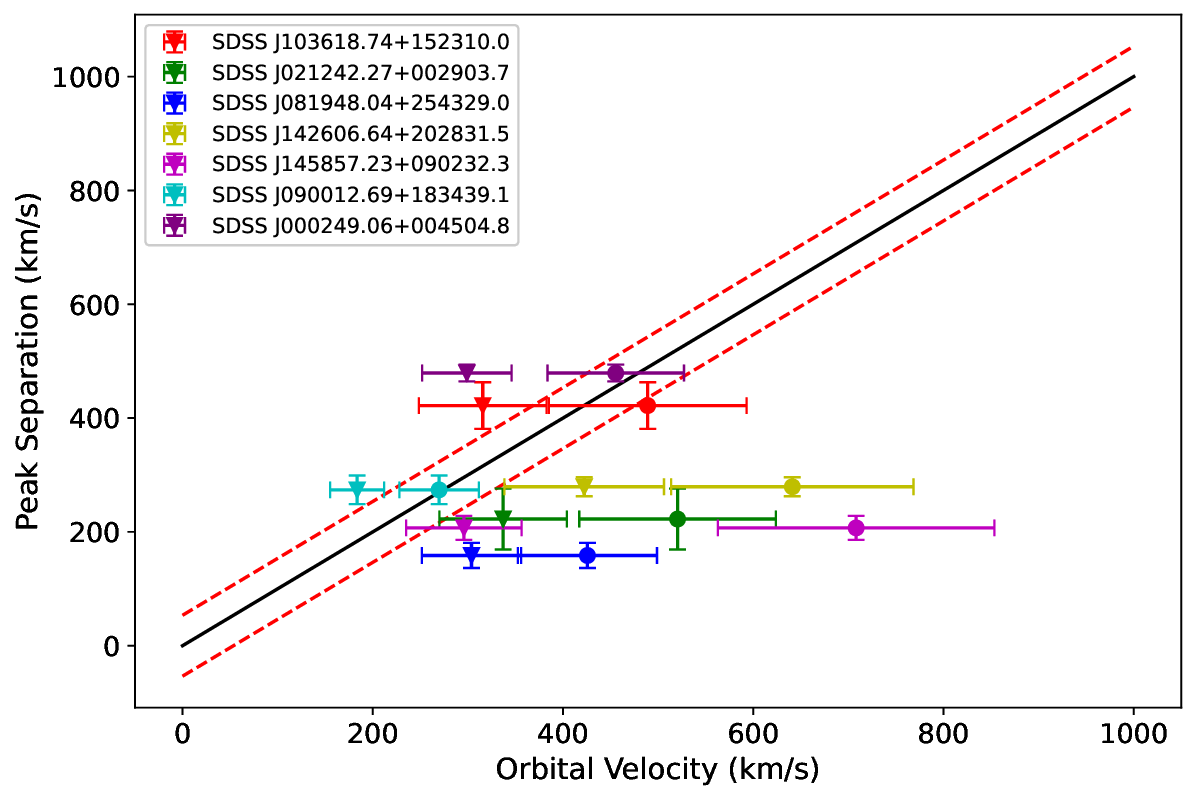}
  \centering
  \caption{Comparison between the peak separation of DPNELs and orbital velocities for the seven kpc-scale dual-core systems. 
The circular symbol shows the comparison between peak separation and $V_{max}^{p}$, while the triangular symbol shows the comparison between peak separation 
and $V_{max}^{c}$. 
Different dual-core systems are presented in different colors, as shown in the legend in the top left corner. 
The solid black line marks the 1:1 relation where peak separation equals to orbital velocity, with red dashed lines indicating uncertainties derived 
from the maximum peak separation errors in our sample.}
  \label{orbi}
\end{figure}

\section{MAIN RESULTS}
Once considering that DPNELs arise from a dual-core system, the maximum orbital velocity ($V_{max}$) should be larger than or equal to the peak 
separation of the DPNELs.
Section 2 provides the total stellar mass and projected distance for each dual-core system. 
Using equation (1) with the assumption of $sin\theta \times cos\beta$=1, the primary $V_{max}$ (without separating the overlapping components, 
hereafter $V_{max}^{p}$) and the corrected $V_{max}$ (with separating the overlapping components, hereafter $V_{max}^{c}$) are measured. 
The uncertainties of the velocities are determined by the mass uncertainties provided in Appendix B.
Table \ref{main_results} lists $V_{max}^{p}$ and $V_{max}^{c}$ for all seven kpc-scale dual-core systems. 
As shown in Figure \ref{orbi}, before the decomposition process, the $V_{max}^{p}$ of the seven systems are larger than or close to their peak 
separations. 

After separating the overlapping components in the photometric images, the peak separations of the two dual-core systems 
separated by $\geq$ 3$\arcsec$ (SDSS J090012.69+183439.1, SDSS J000249.06+004504.8) are larger than their $V_{max}^{c}$. 
Thus, the method proposed in the manuscript verifies that DPNELs in a dual-core system with a larger projected distance ($\geq$ 3$\arcsec$) 
are unlikely produced by orbital kinematics.
The observed DPNELs in such systems may instead arise from extended [O~\textsc{iii}] emission lines in the companion galaxy, 
whose NLR could span $\sim$10 kpc, overlapping with the main galaxy's NLR. 
This leads both NLRs to be detected within a single fiber.

For the five dual-core systems separated by $\leq$ 3$\arcsec$, after the decomposition, SDSS J103618.74+152310.0 exhibits a peak separation 
larger than its $V_{max}^{c}$, indicating that its DPNELs are not due to orbital kinematics. 
The other four systems show peak separations comparable to or smaller than their $V_{max}^{c}$, indicating that their DPNELs can be  
explained by orbital kinematics.
However, the upper limit of the orbital velocity is derived from Equation (1) with assuming sin$\theta$ $\times$ cos$\beta$=1. 
In reality, $\theta$ and $\beta$ are unknown, and other possible values will lead sin$\theta$ $\times$ cos$\beta$ to be smaller than 1. 
Therefore, we evaluate the likelihood that the DPNELs in these four systems are not kinematic in origin, by computing corrected orbital velocities 
($V^{c}$) under varying $\theta$ and $\beta$ and comparing them to the observed peak separations.
Therefore, by giving random $\theta$ and $\beta$ (from -90\textdegree to 90\textdegree), we calculate possible $V^{c}$ with different assumptive 
distances ($D/(sin\theta \times cos\beta$) for 10,000 times to compare with their peak separations.
The results present that the probabilities of peak separation larger than $V^{c}$ with random $\theta$ and $\beta$ for  
the four systems are: 57.11\% (SDSS J021242.27+002903.7), 30.67\% (SDSS J081948.04+254329.0), 
57.76\% (SDSS J142606.64+202831.5), and 62.12\% (SDSS J145857.23+090232.3). 

Before ending this section, several additional points require emphasis.

Although the IMF of \cite{2001MNRAS.322..231K} is preferred for elliptical galaxies in stellar mass estimations, \cite{2003MNRAS.341...33K} 
demonstrates that adopting the IMF of \cite{1955ApJ...121..161S} would increase the stellar mass by a factor of 2. 
Moreover, \cite{2004NewA....9..329P} shows that dynamical masses (accounting for dark matter components) should be 
2$\sim$3 times larger than the stellar masses. 
Considering these effects, our stellar mass estimates may be underestimated by a factor of 4$\sim$6. 
To assess the impact of this systematic bias, we reevaluated the probabilities of peak separations being larger than $V^{c}$ 
(with random $\theta$ and $\beta$) for all seven objects, assuming a 6-fold increase in total stellar mass. 
The recalculated probabilities (in the order of Column 2 in Table \ref{main_results}) are: 43.25\%, 15.04\%, 7.25\%, 14.88\%, 16.25\%, 47.64\%, 
and 56.46\%. 

In this manuscript, $V_{max}$ of a dual-core system is estimated under the assumption of a circular orbit.  
However, if the system follows an elliptical orbit, orbital eccentricity would affect the velocities, with maximum values occurring at periapsis 
and minimum values at apoapsis. 
Consequently, defining an upper limit for the variable orbital velocity in elliptical orbits becomes problematic, given unknown eccentricities 
and uncertain positions of the galaxies along their orbital path. 
Thus, $V_{max}^{c}$ derived for circular orbits may either overestimate or underestimate velocities in elliptical orbital scenarios.

If the two galaxies in a dual-core system were interacting rather than orbiting each other, the interaction process would likely produce 
tidal tails. 
However, no such features are evident in the images of Figure \ref{GALFIT}, indicating that these systems are not currently undergoing interaction. 
Besides, for objects whose DPNLEs exhibit peak separations larger than $V_{max}^{c}$, the biconical outflow can provide a plausible explanation, 
as described in \cite{2011ApJ...735...48S}. 
However, as this manuscript focuses on testing orbital kinematics as the origin of DPNELs, further and detailed discussions on the biconical 
outflow model are beyond the scope of the manuscript.

\section{CONCLUSIONS}
In this manuscript, a method is proposed to test the physical origin of DPNELs from the orbital kinematics of dual-core
systems, through a sample of seven kpc-scale dual-core systems with DPNELs in SDSS. The main conclusions are as follows:
\begin{itemize}
  \item The impact of overlapping components on photometric results can be explored through \textit{GALFIT} decomposition, 
yielding Petrosian magnitudes for each galaxy in each dual-core system both with and without separating the overlapping components. 
  \item Based on the measured correlation between logarithmic total stellar mass and absolute Petrosian magnitude, 
the total stellar masses of both galaxies in each dual-core system can be determined.
  \item The peak separations of DPNELs, clearly measured after subtracting host galaxy starlight via the pPXF code, 
are mainly similar to the reported results in the literature.
  \item After separating the overlapping components, the DPNELs in two dual-core systems separated by $\geq$ 3$\arcsec$ and one 
system separated by $\leq$ 3$\arcsec$ are not due to orbital kinematics, likely originating from biconical outflows or NLRs kinematics.
For the remaining four dual-core systems separated by $\leq$ 3$\arcsec$, their DPNELs can be explained by orbital kinematics. 
However, simulations assuming different orbital angles obtain the probabilities of DPNELs not being caused by orbital kinematics 
for them: 57.11\%, 30.67\%, 57.76\%, and 62.12\%, respectively.
\end{itemize}

\section*{ACKNOWLEGEMENTS}
We gratefully acknowledge the anonymous referee for giving us constructive comments and suggestions to greatly improve our paper.
Zhang gratefully acknowledges the kind financial support from GuangXi University, and the grant support from NSFC-12173020 and 
NSFC-12373014.
Chen $\&$ Cheng gratefully acknowledge the kind grant support from Innovation Project of Guangxi Graduate Education YCSW2024006.
This manuscript has made use of the data from SDSS projects. The SDSS-III website is \url{http://www.sdss3.org/}. 
The SDSS DR16 website is \url{http://skyserver.sdss.org/dr16/en/home.aspx}. 
We have made use of the data from MILES library (\url{http://miles.iac.es/}) developed for stellar population synthesis models.
This manuscript has made use of \textit{GALFIT} \url{https://users.obs.carnegiescience.edu/peng/work/galfit/galfit.html}, 
the pPXF code \url{http://www-astro.physics.ox.ac.uk/~mxc/idl/#ppxf}, 
the MPFIT package \url{https://pages.physics.wisc.edu/~craigm/idl/cmpfit.html}, 
and the \texttt{lts\_linefit} code \url{https://www-astro.physics.ox.ac.uk/~cappellari/idl/#lts}.

\section*{Data Availability}
The data underlying this article will be shared on reasonable request to the corresponding author
(\href{mailto:xgzhang@gxu.edu.cn}{xgzhang@gxu.edu.cn}).

\newpage
\appendix

\section{}
\begin{figure*}
  \centering
  \includegraphics[width=0.73\paperwidth]{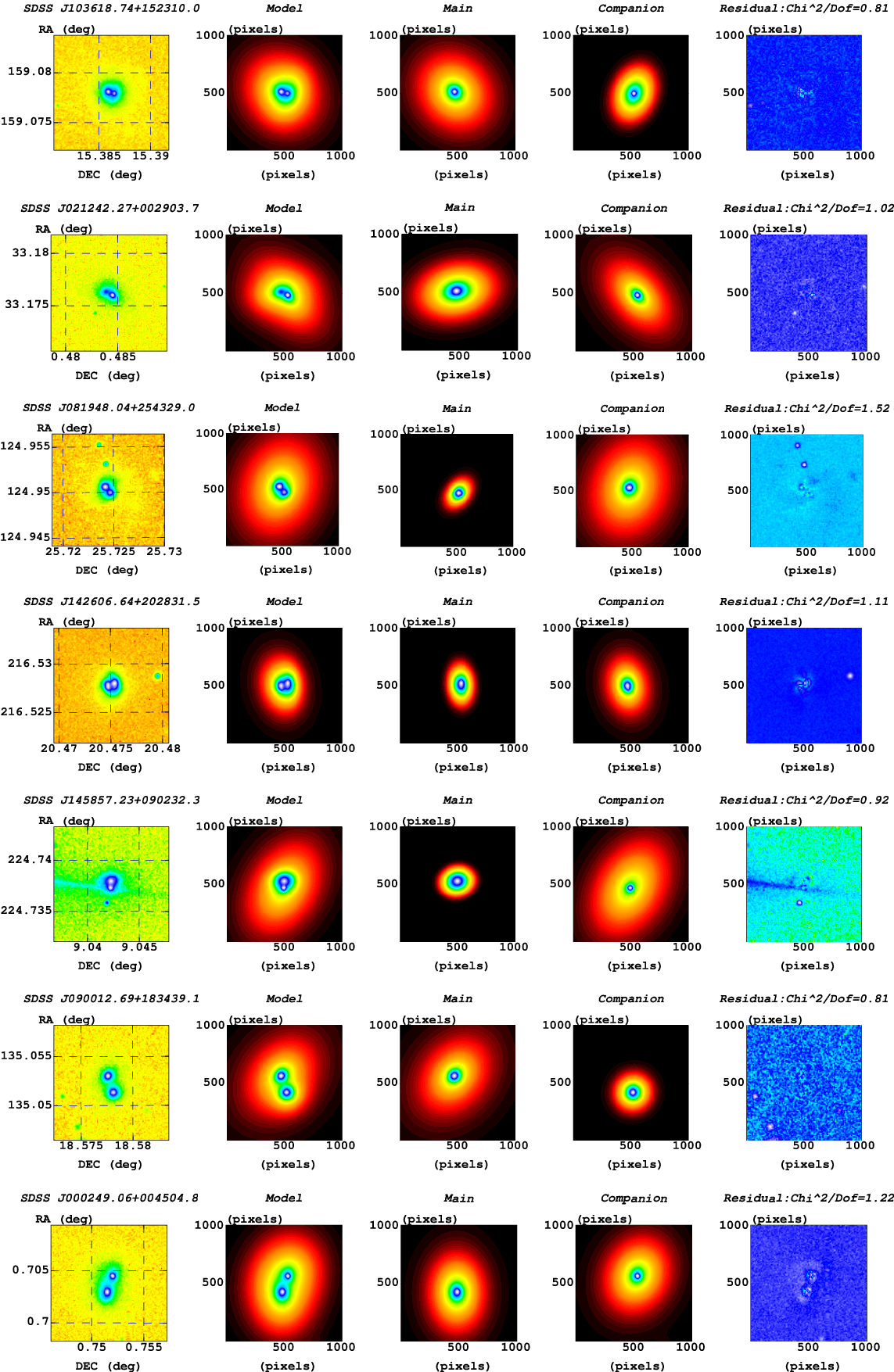}
  \caption{The best fitting results by \textit{GALFTI} to the $r$ band photometric images of the seven kpc-scale dual-core systems. 
Column 1: The image region cut from the FITS image of the SDSS field, with dashed lines 
representing the coordinates.
Column 2: The best fitting results. 
Column 3: The separated image of the main galaxy determined by \textit{GALFIT}. 
Column 4: The separated image of the companion galaxy determined by \textit{GALFIT}. 
Column 5: The residual image with the corresponding $\chi2/dof$ marked in the title.
All images are expanded from 101$\times$101 pixels to 1010$\times$1010 pixels using linear interpolation for enhanced visual clarity.}
  \label{GALFIT_all}
\end{figure*}

The best fitting results determined by \textit{GALFIT} for the seven kpc-scale dual-core systems are shown in Figure \ref{GALFIT_all}, 
including separated images of the main and companion galaxies in each system. 

\begin{figure*}
  \centering
  \includegraphics[width=0.82\paperwidth]{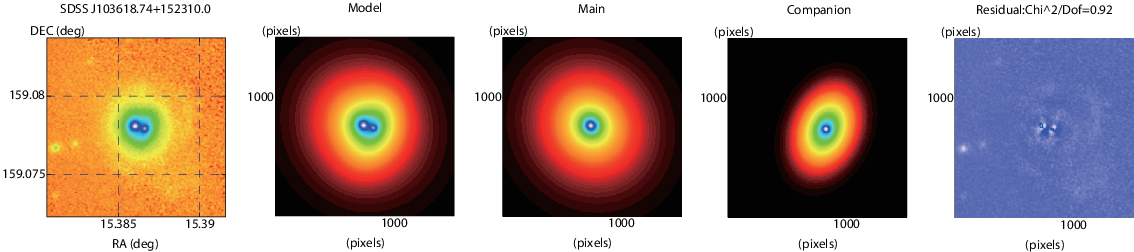}
  \caption{The best fitting results by \textit{GALFIT} for the $r$ band DESI image of the dual-core system in SDSS J103618.74+152310.0. 
Column 1: The image region cut from the FITS image of the DESI field, with dashed lines representing the coordinates.
Column 2: The best fitting results. 
Column 3: The separated image of the main galaxy determined by \textit{GALFIT}. 
Column 4: The separated image of the companion galaxy determined by \textit{GALFIT}. 
Column 5: The residual image with the corresponding $\chi2/dof$ marked in the title.
}
  \label{desi_image}
\end{figure*}

As discussed in \cite{2006AJ....131.2018B, 2020ApJ...899..154S}, the ability to distinguish two cores depends on image qualities.  
Since PanSTARS (pixelscale=0.25 arcsec/pixel) and DESI (pixelscale=0.262 arcsec/pixel) provide higher-quality images than SDSS 
(pixelscale=0.396 arcsec/pixel), we also perform the decomposition on a DESI image of the dual-core system in SDSS J103618.74+152310.0 
using \textit{GALFIT} as a test case. 
The photometric results in Figure \ref{desi_image} show that the resolved images closely match those from SDSS. 

We further measure the related photometric parameters in the separated DESI images of this dual-core system. 
The apparent Petrosian magnitudes of the main and companion galaxies are 16.34 and 16.89, slightly differing from the SDSS-derived values 
of 16.39 and 16.99.  
The magnitude differences between the two galaxies measure 0.6 in SDSS and 0.55 in DESI. 
The Petrosian radii are 7.24\arcsec\ and 6.18\arcsec in DESI, compared to 8.59\arcsec and 6.08\arcsec in SDSS. 
The S{\'e}rsic index (determined by \textit{GALFIT}) are 3.72 and 1.04 for the main and companion galaxies in DESI, while their 
$IC$ are 0.27 and 0.31. 
Similarly, SDSS measurements yield S{\'e}rsic index of 3.26 and 0.94, and $IC$ values of 0.25 and 0.20.
In summary, except for a larger difference of 0.11 in the companion galaxy's $IC$ between DESI and SDSS, most parameters show close agreement, 
supporting that our separating procedure is efficient enough. 
However, due to the SDSS photometric results mainly considered in this manuscript, further discussions on the differences between the SDSS and 
DESI images are beyond the scope of the manuscript.

\section{}
Appendix B provided additional results for the seven kpc-scale dual-core systems, including logarithmic total stellar masses and apparent Petrosian 
magnitudes measured in the images of $g,r,i$ bands, as well as final primary/corrected logarithmic total stellar masses with corresponding uncertainties. 
Here, the uncertainties of the final masses (Column 9 and Column 16 of Table \ref{mass_mag_gri}) are calculated by scaling the ratio of  
average mass uncertainty to average mass for the 901 elliptical galaxies in the mass-magnitude sample.

Moreover, the offsets in arcseconds between the drilled positions of SDSS fibers (shown by plug$\_$ra and plug$\_$dec in SDSS) and the 
\textit{GALFIT} determined central positions are provided. 
As listed in Column 17 of Table \ref{mass_mag_gri}, all offsets are smaller than 0.64$\arcsec$, indicating that SDSS spectra adequately cover the 
central regions of the galaxies.

\begin{table*}
\begin{threeparttable}
 \caption{Additional Results}
 \label{mass_mag_gri}
 
 \begin{tabular*}{0.88\paperwidth}{c@{\hspace*{6pt}}c@{\hspace*{6pt}}c@{\hspace*{6pt}}c@{\hspace*{6pt}}c@{\hspace*{5pt}}c@{\hspace*{5pt}}c
@{\hspace*{5pt}}c@{\hspace*{5pt}}c@{\hspace*{5pt}}c@{\hspace*{5pt}}c@{\hspace*{5pt}}c@{\hspace*{5pt}}c@{\hspace*{5pt}}c@{\hspace*{5pt}}c
@{\hspace*{5pt}}c@{\hspace*{5pt}}c@{\hspace*{5pt}}c}
 
 \hline
  Type & SDSS Name & $m_{p}^{g}$ & $m_{p}^{r}$ & $m_{p}^{i}$ & $M_{*}^{p}(g)$ & $M_{*}^{p}(r)$ & $M_{*}^{p}(i)$ & $M_{*}^{p}$
& $m_{c}^{g}$ & $m_{c}^{r}$ & $m_{c}^{i}$ & $M_{*}^{c}(g)$ & $M_{*}^{c}(r)$ & $M_{*}^{c}(i)$ & $M_{*}^{c}$ & $\triangle D$\\
  (1) & (2) & (3) & (4) & (5) & (6) & (7) & (8) & (9) & (10) & (11) & (12) & (13) & (14) & (15 & (16)) & (17)\\
 \hline
 main    & J103618.74+152310.0 & 16.77 & 15.87 & 15.39 & 10.96 & 10.96 & 10.98 & 10.97±0.12 & 18.01 & 16.39 & 16.00 & 10.35 & 10.72 & 10.70 & 10.59±0.12 & 0.40\\
 comp    & J103618.65+152311.9 & 16.88 & 15.96 & 15.47 & 10.92 & 10.93 & 10.96 & 10.94±0.12 & 17.37 & 16.99 & 16.39 & 10.68 & 10.45 & 10.54 & 10.56±0.12 & 0.51\\
\hline
 main    & J021242.27+002903.7 & 17.71 & 16.61 & 16.14 & 11.43 & 11.51 & 11.51 & 11.48±0.12 & 18.54 & 17.58 & 17.24 & 11.02 & 11.05 & 11.01 & 11.03±0.12 & 0.25\\
 comp    & J021242.18+002906.1 & 17.78 & 16.71 & 16.25 & 11.39 & 11.45 & 11.46 & 11.43±0.12 & 18.55 & 17.32 & 16.80 & 11.01 & 11.16 & 11.20 & 11.12±0.12 & 0.49\\
\hline
 main    & J081948.04+254329.0 & 17.13 & 16.26 & 15.80 & 11.02 & 11.01 & 11.02 & 11.02±0.12 & 18.47 & 17.48 & 16.76 & 10.37 & 10.44 & 10.58 & 10.46±0.12 & 0.64\\
 comp    & J081948.22+254330.4 & 16.52 & 15.71 & 15.21 & 11.32 & 11.27 & 11.29 & 11.30±0.12 & 16.84 & 16.01 & 15.85 & 11.17 & 11.13 & 11.00 & 11.10±0.12 & 0.44\\
\hline
 main    & J142606.64+202831.5 & 16.26 & 15.51 & 15.06 & 11.38 & 11.29 & 11.29 & 11.32±0.12 & 16.90 & 16.20 & 15.79 & 11.07 & 10.97 & 10.96 & 11.00±0.12 & 0.23\\
 comp    & J142606.51+202829.8 & 16.26 & 15.51 & 15.04 & 11.38 & 11.29 & 11.31 & 11.33±0.12 & 17.19 & 16.35 & 15.82 & 10.92 & 10.90 & 10.95 & 10.92±0.12 & 0.22\\
\hline
 main    & J145857.23+090232.3 & 17.36 & 16.75 & 16.30 & 11.51 & 11.35 & 11.35 & 11.40±0.12 & 18.40 & 17.30 & 16.91 & 10.99 & 11.09 & 11.07 & 11.05±0.12 & 0.31\\
 comp    & J145857.08+090232.7 & 16.13 & 16.14 & 15.51 & 12.10 & 11.62 & 11.70 & 11.81±0.12 & 19.50 & 18.26 & 17.33 & 10.44 & 10.63 & 10.86 & 10.65±0.12 & 0.38\\
\hline
 main    & J090012.69+183439.1 & 17.10 & 16.29 & 15.86 & 11.03 & 10.98 & 10.98 & 11.00±0.12 & 17.80 & 16.95 & 16.49 & 10.68 & 10.67 & 10.69 & 10.68±0.12 & 0.45\\
 comp    & J090012.30+183436.8 & 17.26 & 16.47 & 16.02 & 10.93 & 10.89 & 10.89 & 10.90±0.12 & 18.00 & 17.21 & 16.81 & 10.57 & 10.54 & 10.53 & 10.54±0.12 & 0.47\\
\hline
 main    & J000249.06+004504.8 & 16.55 & 15.59 & 15.13 & 11.38 & 11.39 & 11.39 & 11.39±0.12 & 17.24 & 16.23 & 15.73 & 11.04 & 11.09 & 11.12 & 11.08±0.12 & 0.59\\
 comp    & J000249.43+004506.7 & 16.47 & 15.52 & 15.04 & 11.42 & 11.42 & 11.43 & 11.42±0.12 & 17.35 & 16.44 & 15.97 & 10.98 & 10.99 & 11.01 & 10.99±0.12 & 0.46\\

\hline
 \end{tabular*}
\begin{tablenotes}[flushleft]
\setlength\labelsep{0pt}
\small
\item{Column (1) shows the component of a dual-core system, main represents the main galaxy with DPNELs, comp represents the companion galaxy; 
Column (2) shows the names of the galaxies; 
Column (3), (4) and (5) show the primary apparent Petrosian magnitudes measured from the images of $g,r,i$ bands; 
Column (6), (7) and (8) show the primary logarithmic total stellar masses of $g,r,i$ bands, 
Column(9) shows the final primary logarithmic total stellar mass measured by averaging the primary masses of $g,r,i$ bands; 
Column (10), (11) and (12) show the corrected apparent Petrosian magnitudes measured from the images of $g,r,i$ bands; 
Column (13), (14) and (15) show the corrected logarithmic total stellar masses of $g,r,i$ bands, 
Column(16) shows the final corrected logarithmic total stellar mass measured by averaging the corrected masses of $g,r,i$ bands;
Column (17) shows the offsets in arcseconds between the drilled positions of SDSS fibers and the \textit{GALFIT} determined central positions.  
}
\end{tablenotes}
\end{threeparttable}
\end{table*}

\section{}
\begin{figure*}
  \centering
  \includegraphics[width=0.8\paperwidth]{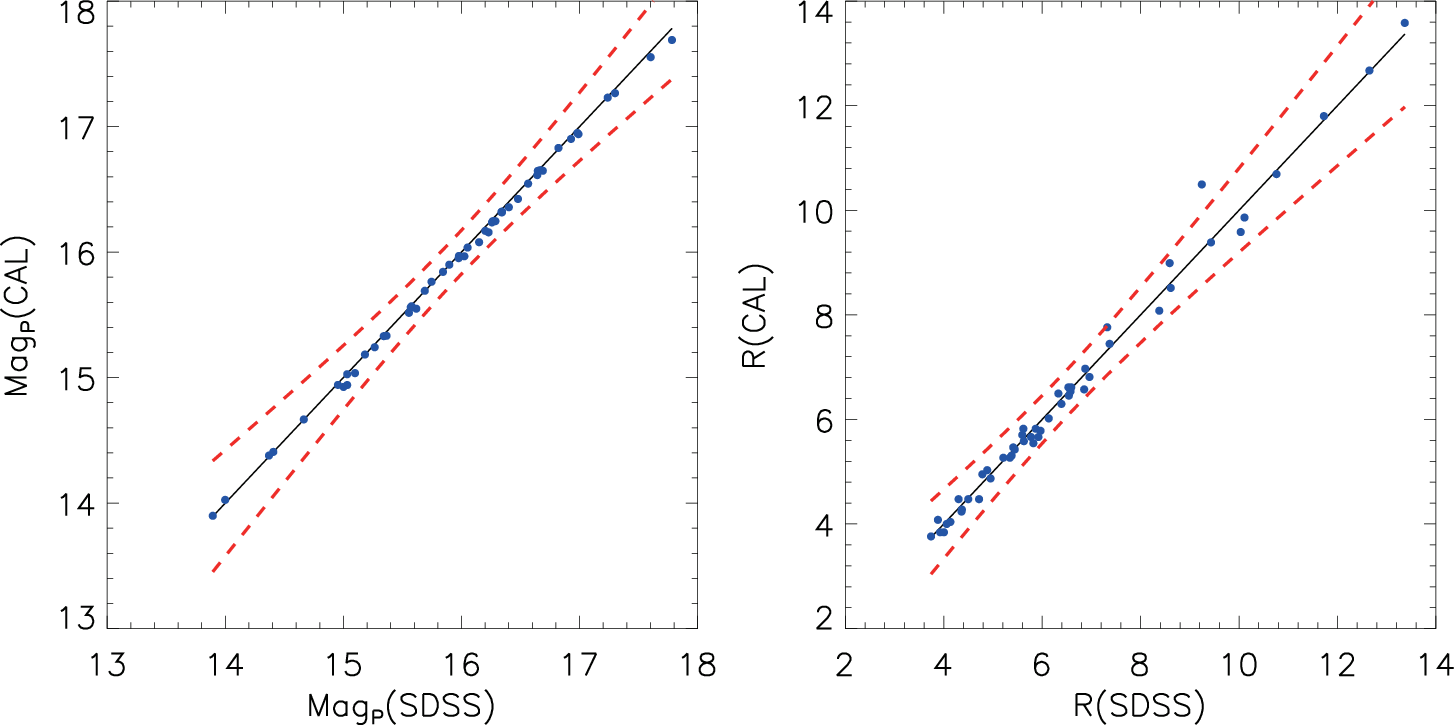}
  \caption{On the correlation between the SDSS provided Petrosian parameters and recalculated values for the 50 randomly selected single galaxies. 
The solid circles in blue represent the values of apparent Petrosian magnitudes (left panel) and Petrosian radii (right panel). 
In both the left and right panels, the solid lines in black show Y=X, and the dashed lines in red show the corresponding 5$\sigma$ confidence bands.}
  \label{single_verify}
\end{figure*}

\begin{figure*}
  \centering
  \includegraphics[width=0.8\paperwidth]{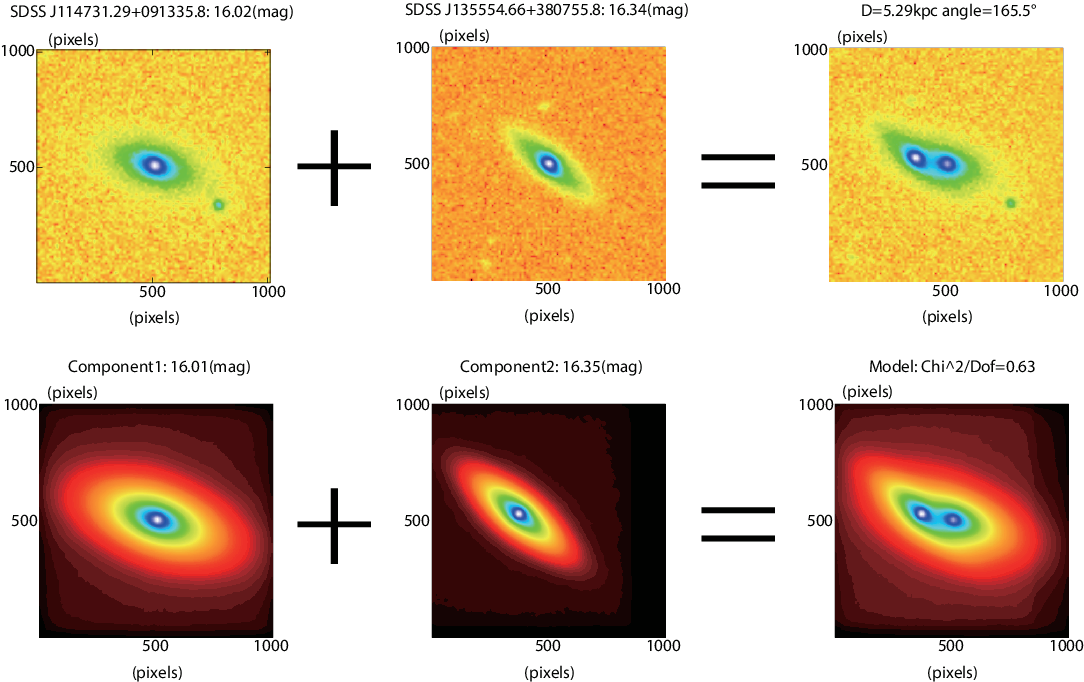}
  \caption{An example of the artificial dual-core system image. 
The top left and top middle panels show the original photometric images of the two randomly selected galaxies, the name and $r$ band apparent Petrosian magnitude provided by SDSS for each galaxy are shown in the title of each panel. 
The top right panel shows the overlapped image with a projected distance of 5.29 kpc and an angle of 165.5°.
The bottom left and bottom middle panels show the separated images determined by applying \textit{GALFIT} to the overlapped image shown in the top right panel, the recalculated apparent Petrosian magnitude of each galaxy is shown in the title of each panel. 
The bottom right panel shows the best fitting results determined by \textit{GALFIT} with the corresponding $\chi2/dof$=0.63.}
  \label{double_verify}
\end{figure*}

\begin{figure}
  \centering
  \includegraphics[width=\columnwidth]{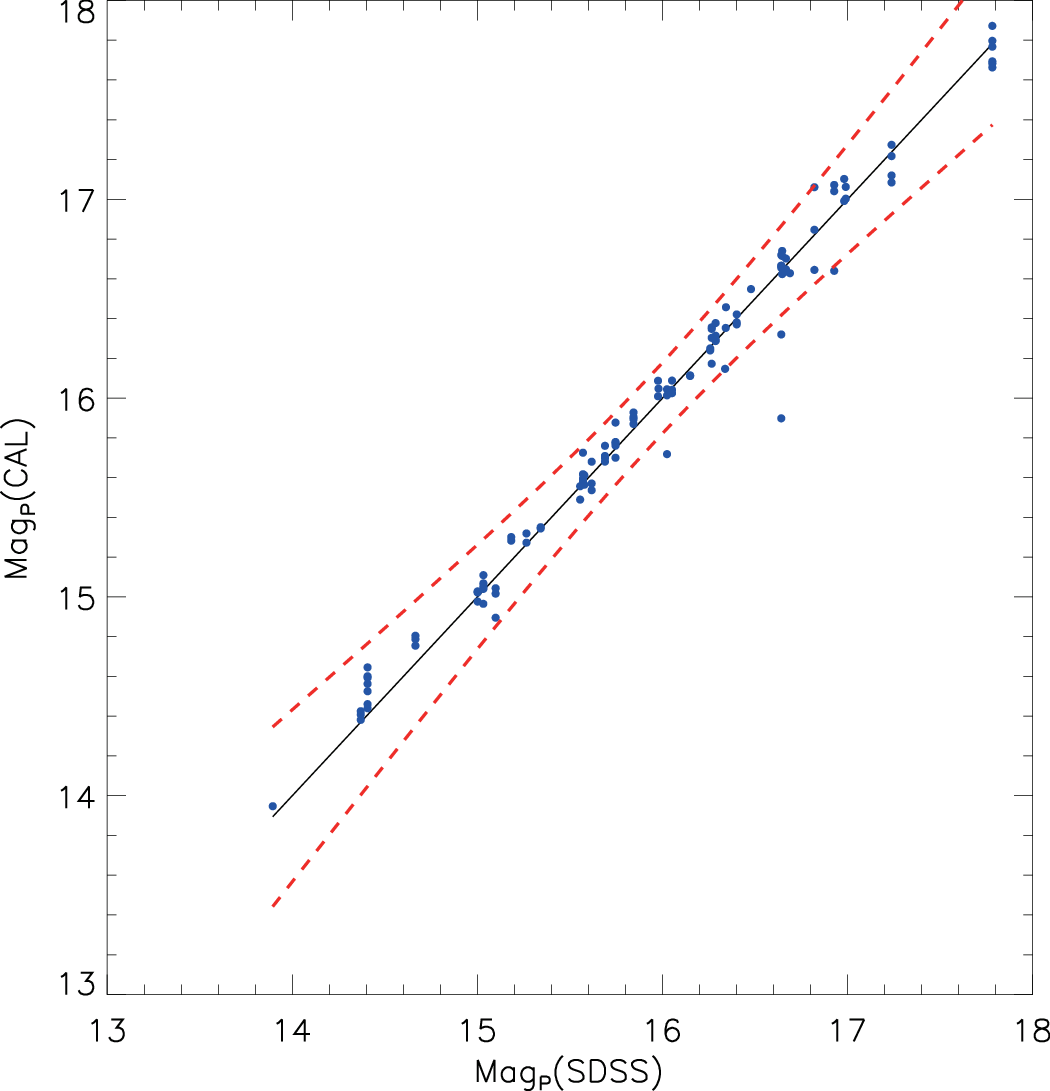}
  \caption{On the correlation between the SDSS provided apparent Petrosian magnitudes and recalculated values derived from the the separated images 
for the artificial dual-core system galaxies. The black solid line shows Y=X, and the red dashed lines show the corresponding 5$\sigma$ confidence bands.}
  \label{double_verify_mag}
\end{figure}

To validate the accuracy of recalculated Petrosian parameters, 50 single galaxies are randomly selected from the mass-magnitude sample for analysis.  
As shown in Figure \ref{single_verify}, the recalculated apparent Petrosian magnitudes and Petrosian radii show no significant deviation 
from SDSS provided values. 

To assess degeneracy effects on SDSS images that might lead to nonunique decompositions of dual-core systems, we generate 60 artificial 
dual-core system images by overlapping pairs of galaxies randomly selected from the above 50-sample, with random orientation angles 
(relative to the direction of the horizontal x-axis) and projected distances spanning 2.5 $\sim$ 5.5 kpc -- the minimum and maximum separations 
observed in our seven kpc-scale dual-core systems. 
After applying \textit{GALFIT} to separate the artificial overlapping images, we recalculate apparent Petrosian magnitudes from the separated images. 
Figure \ref{double_verify} shows an example of the artificial dual-core system image with the best fitting results as well as separated images 
determined by \textit{GALFIT}. 
Figure \ref{double_verify_mag} demonstrates that the recalculated apparent Petrosian magnitudes closely match SDSS values, indicating 
that there are few effects of degeneracies on the separating procedure. 

\section{}
\begin{figure*}
  \includegraphics[width=0.65\paperwidth]{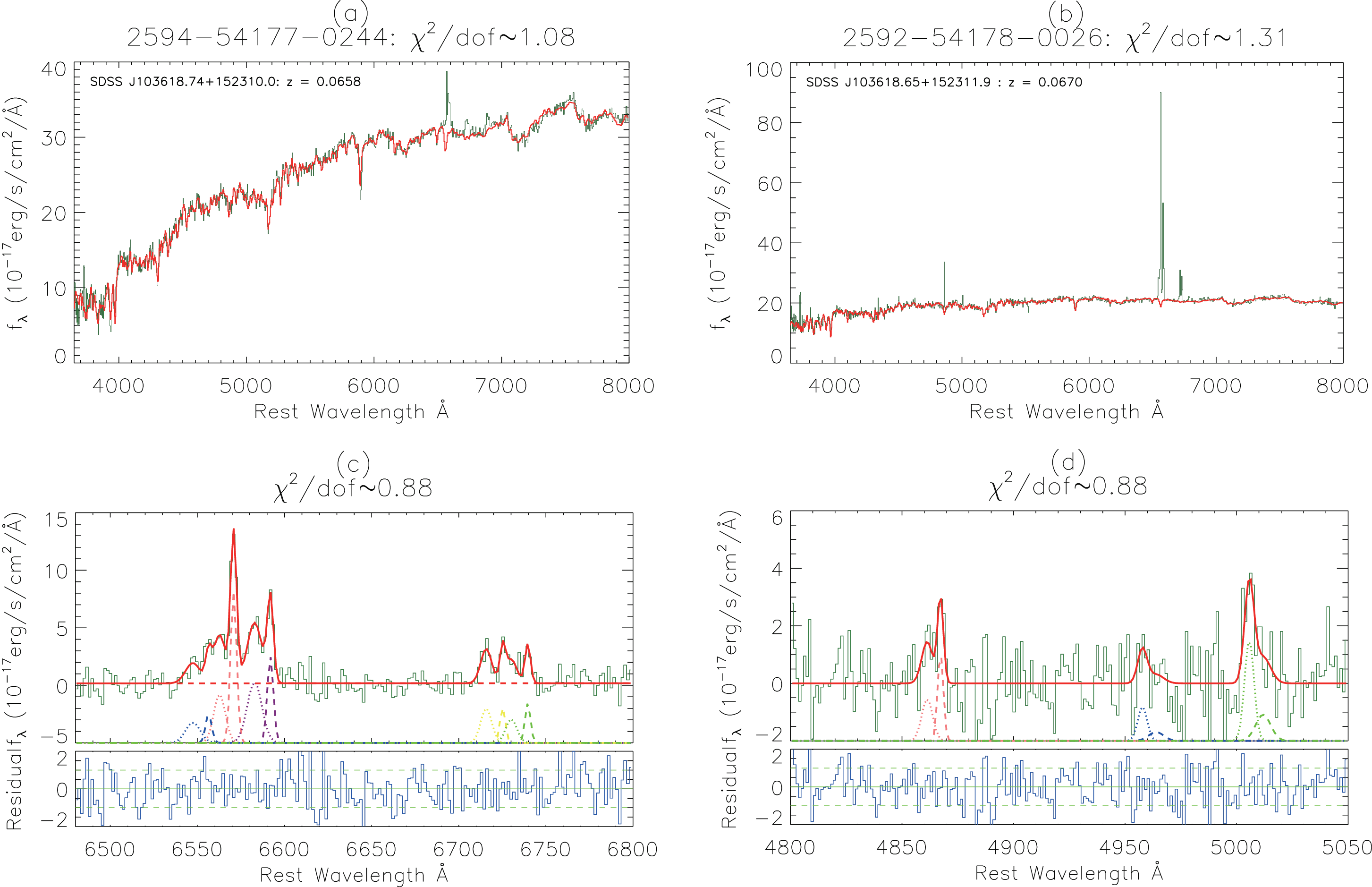}
  \caption{ Panel (a): the rest-frame spectroscopic results of the main galaxy with DPNELs. 
Panel (b): the rest-frame spectroscopic results of the companion galaxy with single-peaked narrow emission lines. 
In panels (a) and (b), the lines in dark green show the SDSS spectra, and the lines in red show the host galaxy starlight determined 
by the pPXF code. 
Panel (c): the best fitting results to the emission lines around H$\alpha$ in the main galaxy spectrum shown in the top left panel.
Panel (d): the best fitting results to the emission lines around H$\beta$ in the main galaxy spectrum shown in the top left panel.
In panels (c) and (d), the solid lines in dark green represent the spectra by subtracting the starlight of host galaxy; the solid red 
lines represent the best fitting results for the emission lines; the solid blue lines in the bottom regions show the residuals; 
the long and short dashed lines in purple, in pink, in blue, in yellow, and in green for the double-peak features in 
[N~\textsc{ii}]$\lambda$ 6550\AA, H$\alpha$, [N~\textsc{ii}]$\lambda$ 6585\AA, [S~\textsc{ii}]$\lambda$ 6718\AA, 
and [S~\textsc{ii}]$\lambda$ 6732\AA\hspace{0.05cm} narrow emission lines in panel (c); 
the long and short dashed lines in pink, in blue, and in green for the double-peak features in H$\beta$, 
[O~\textsc{iii}]$\lambda$ 4959\AA, and [O~\textsc{iii}]$\lambda$ 5007\AA\hspace{0.05cm} narrow emission lines in panel (d). 
In the title of each panel, the corresponding $\chi2/dof$ related to the best fitting results are marked.}
  \label{spectroscopic}
\end{figure*}

Figure \ref{spectroscopic} shows the spectroscopic results of the dual-core system in SDSS J103618.74+152310.0.

\label{lastpage}
\end{document}